\documentclass{moriond}

\def\be{\begin{equation}}
\def\ee{\end{equation}}
\def\bea{\begin{eqnarray}}
\def\eea{\end{eqnarray}}

\newcommand{\Photo}{}

\begin{document}
\vspace*{4cm}
\title{ANOMALOUS DIMUON CHARGE ASYMMETRY IN $p \bar{p}$ COLLISIONS}

\author{ B. HOENEISEN }

\address{(D\O\ Collaboration) \\
Universidad San Francisco de Quito, Quito, Ecuador}

\maketitle\abstracts{
We present an overview of the measurements of the like-sign
dimuon charge asymmetry by the D\O\ Collaboration 
at the Fermilab Tevatron $p \bar{p}$
Collider. The results differ from the Standard Model
prediction of CP violation in mixing and interference
of $B^0$ and $B^0_s$ by 3.6 standard deviations.
}

\section{Introduction}
\begin{figure}
\begin{center}
\scalebox{0.8}
{\includegraphics{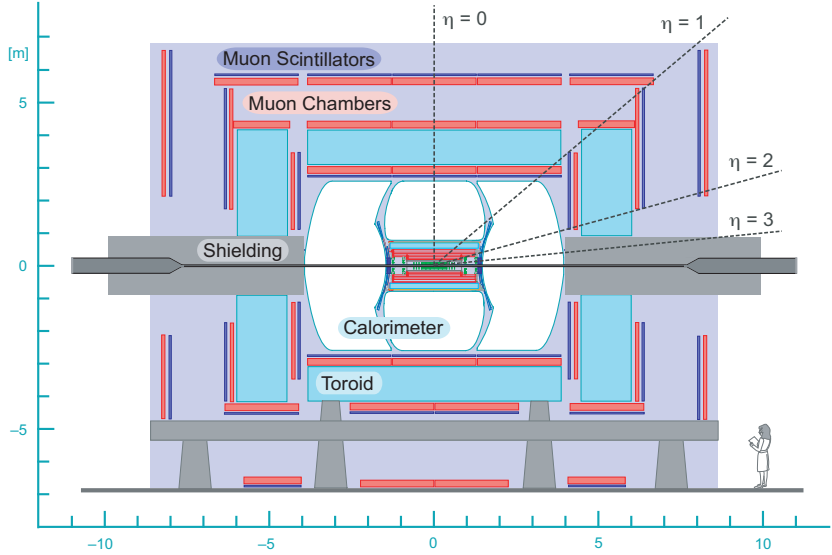}}
\caption{The D\O\ detector.}
\end{center}
\label{d0}
\end{figure}

The D\O\ Collaboration has measured the inclusive muon charge asymmetry
and the like-sign dimuon charge asymmetry 
in $p \bar{p}$ collisions at 
a center of mass energy $\sqrt{s} = 1.96$ TeV
at the Fermilab Tevatron 
Collider \cite{dimu_asym1}$^,$ \cite{dimu_asym2}$^,$ 
\cite{dimu_asym3}$^,$ \cite{dimu_asym4}.
The measured inclusive muon charge asymmetry is consistent with
zero, while the like-sign dimuon charge asymmetry is significantly negative.
The result differs from the Standard Model 
prediction of CP violation in mixing, and CP violation in interference
of decay amplitudes with and without mixing, of $B^0$ and
$B^0_s$ by 3.6 standard deviations \cite{dimu_asym4}. In this
note we present an overview of these measurements and their current
status.
The D\O\ detector is shown in Figure \ref{d0}.

The original motivation for this measurement was
CP violation in mixing \cite{Branco}.
Most like-sign dimuons at D\O\ (after removing background
muons from kaon and pion decay) are from events
with a $b \bar{b}$ pair. One of the $b$'s decays to a
``right sign" muon, i.e. to a muon of the same charge
sign as the parent $b$ quark, while the other $b$ in
the event hadronizes to a $B^0$ or $B^0_s$ meson that
oscillates before decaying to a ``wrong sign" muon.
CP violation is expected to be negligible for ``right
sign" muons, while it is expected to be non-negligible
for ``wrong sign" muons due to CP violation in mixing \cite{Branco}
and, as was recently realized \cite{CPV_interference},
also due to CP violation in interference of decay amplitudes
with and without mixing, of
$B^0$ and $B^0_s$.
The charge asymmetry of inclusive muons is predicted
to be very small compared to the present experimental
uncertainty \cite{dimu_asym4} because most muons of events with a
$b \bar{b}$ or a $c \bar{c}$ are ``right sign" muons,
and because CP violation in interference does not contribute.
Therefore, the inclusive muon charge asymmetry provides a null-test
that validates the measurements of the detector and background
charge asymmetries. 
Like-sign dimuons have one ``right sign" muon which tags the other muon
to be of the ``wrong sign". 

\section{Definitions}
Most $p \bar{p}$ collisions produce no muons, a few produce one recorded muon,
very few produce two recorded muons, and fewer still produce two \textit{like-sign}
recorded muons, i.e. $4 \times 10^{14}$ (mostly not recorded), $2.2 \times 10^9$, $ 2.8 \times 10^7$ and
$6.2 \times 10^6$ respectively in the final data set of 10.4 fb$^{-1}$.
We obtain the \textit{raw} inclusive muon and like-sign dimuon charge asymmetries by
counting inclusive muons or like-sign dimuon events:
\begin{equation}
a = \frac{n(\mu^+) - n(\mu^-)}{n(\mu^+) + n(\mu^-)}, \qquad
A = \frac{N(\mu^+ \mu^+) - N(\mu^- \mu^-)}{N(\mu^+ \mu^+) + N(\mu^- \mu^-)}.
\end{equation}
We obtain the \textit{residual} charge asymmetries by
subtracting the asymmetry of the detector and backgrounds (``muons" from charged
kaon and pion decay, and hadrons that punch-through to the outer muon detectors):
\begin{equation}
a_{CP} = a - a_{\textrm{bkg}}, \qquad A_{CP} = A - A_{\textrm{bkg}}.
\end{equation}
These CP-violating residual charge asymmetries $a_{CP}$ and $A_{CP}$
are normalized to all muons or like-sign
dimuons, including background ``muons". The corresponding charge
asymmetries that exclude events with background ``muons" from the normalization are
$a_S$ and $A_S$.
To cancel first order detector asymmetries we collect equal numbers of events for
each of the four solenoid and toroid polarity combinations \cite {dimu_asym1}.
The residual detector asymmetry and the background asymmetries 
are \textit{measured} by reconstructing exclusive 
decays in the \textit{same} data sets \cite{dimu_asym2}$^,$ 
\cite{dimu_asym3}$^,$ \cite{dimu_asym4}. 

\section{History}
The history of the measurements of the like-sign dimuon charge asymmetry at D\O\
is summarized in Table \ref{history}. Note that the results have not changed significantly
with a factor 10 increase of the data set, and many improvements of the analysis 
methods, over the years. The decrease of $|A_{CP}|$ in the last measurement is
due to a finer binning of the measurements in transverse momentum $p_T$ and pseudo-rapidity
$\eta$. The studies, measurements and cross-checks are already 22 years old.

\begin{table}
\caption[]{Measurements of the like-sign dimuon charge asymmetry at D\O.
$\epsilon$ is the discrepancy between the measurement and the Standard Model
prediction for CP violation in mixing (* and interference) of $B^0$ and $B^0_s$.}
\label{history}
\vspace{0.4cm}
\begin{center}
\begin{tabular}{|rccr|}
\hline
$\int L dt$ & Asymmetry $A_{CP}$ & $\epsilon$ & D\O\ , Phys.Rev. D \\
\hline
1.0 fb$^{-1}$  & $(-0.28 \pm 0.13 \pm 0.09)\%$ & 1.7$\sigma$ & {\bf 74}, 092001 (2006) \\
6.1 fb$^{-1}$  & $(-0.252 \pm 0.088 \pm 0.092)\%$ & 3.2$\sigma$ & {\bf 82}, 032001 (2010) \\
9.0 fb$^{-1}$  & $(-0.276 \pm 0.067 \pm 0.063)\%$ & 3.9$\sigma$ & {\bf 84}, 052007 (2011) \\
10.4 fb$^{-1}$ & $(-0.235 \pm 0.064 \pm 0.055)\%$ & 3.6$\sigma$ * & {\bf 89}, 012002 (2014) \\
\hline
\end{tabular}
\end{center}
\end{table}

\section{Measurement with the full data set of 10.4 fb$^{-1}$}
\begin{figure}
\begin{center}
\scalebox{0.4}
{\includegraphics{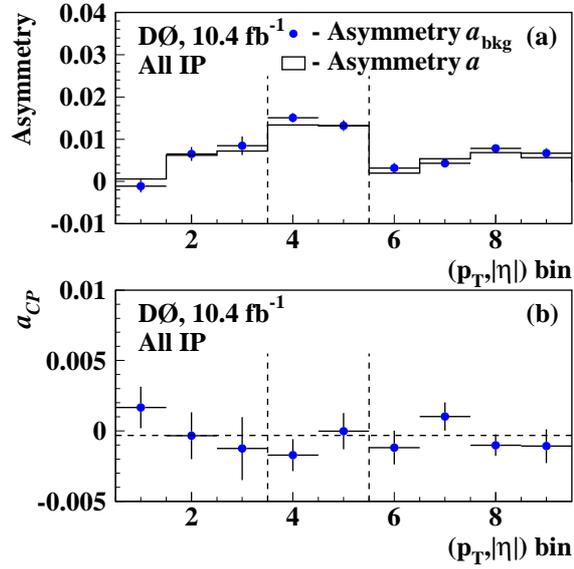}}
\caption{Raw inclusive muon charge asymmetry $a$ obtained by counting muons,
measured detector and background charge asymmetries $a_\textrm{bkg}$,
and their difference $a_{CP} = a - a_\textrm{bkg}$
in 9 bins of $(p_T, |\eta|)$. The horizontal line shows the weighted average
which is consistent with zero.}
\label{a}
\end{center}
\end{figure}

All measurements are done in 9 bins of $(p_T, |\eta|)$ \cite{dimu_asym4}. 
The results for inclusive muons are presented in Figure \ref{a}.
Note that the measured background charge asymmetry $a_\textrm{bkg}$ is consistent with the
raw asymmetry $a$ obtained by counting events in all bins.
We obtain a weighted average
$a_{CP} = (-0.032 \pm 0.042 \textrm{(stat)} \pm 0.061 \textrm{(syst)})\%$, 
which is consistent with zero as expected
(the Standard Model prediction is $a_{CP} = (-0.0007 \pm 0.0002)\%$).
Figure \ref{a} demonstrates that we understand the detector and
background asymmetries within the quoted uncertainties.


\begin{figure}
\begin{center}
\scalebox{0.4}
{\includegraphics{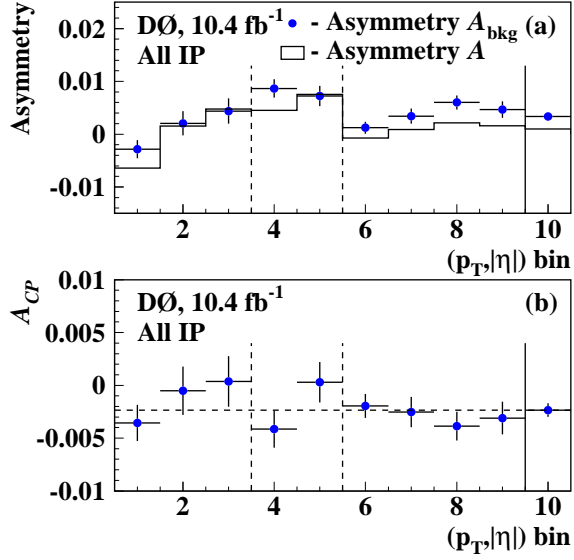}}
\caption{Raw like-sign dimuon charge asymmetry $A$ obtained by counting 
like-sign dimuons,
measured detector and background asymmetries $A_\textrm{bkg}$,
and their difference $A_{CP} = A - A_\textrm{bkg}$
in 9 bins of $(p_T, |\eta|)$. The histogram has two entries per event.
The 10th bin and the horizontal line present 
the weighted average which is significantly negative.}
\label{A}
\end{center}
\end{figure}

The results for like-sign dimuons are presented in Figure \ref{A}.
Note that the measured background charge asymmetry $A_\textrm{bkg}$ follows the
raw asymmetry $A$ obtained by counting events, but there is an 
offset: $A_{CP} = A - A_\textrm{bkg}$ is significantly negative, and is negative
(with limited significance) for the central muon detector (bins 1, 2 and 3), 
forward muon detector (bins 6 to 9), and for
their overlap region (bins 4 and 5). We mention that 
the central and forward muon detectors have different technologies
and their reconstruction softwares are independent.
$A_{CP}$ does not depend significantly on the 
$(p_T, |\eta|)$ bin. The weighted average is
$A_{CP} = (-0.235 \pm 0.064 \textrm{(stat)} \pm 0.055 \textrm{(syst)})\%$,
while the Standard Model prediction is
$A_{CP} = (-0.043 \pm 0.010)\%$. 

\begin{figure}
\begin{center}
\scalebox{0.45}
{\includegraphics{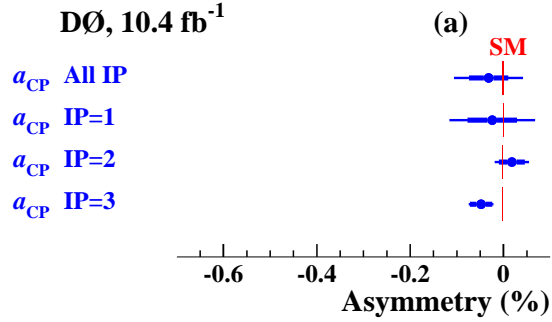}}
\caption{Inclusive muon charge asymmetry $a_{CP}$
for all muons, and for muons with transverse 
impact parameter less than 50 $\mu$m (IP=1),
in the range 50 $\mu$m to 120 $\mu$m (IP=2), and
in the range 120 $\mu$m to 3000 $\mu$m (IP=3).}
\label{aCP}
\end{center}
\end{figure}

\begin{figure}
\begin{center}
\scalebox{0.45}
{\includegraphics{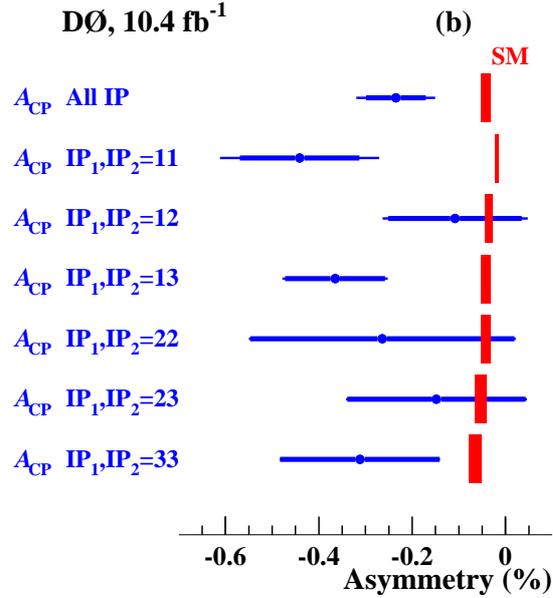}}
\caption{Like-sign dimuon charge asymmetry $A_{CP}$ 
for all muons, and for dimuons in six bins
of (IP$_1$, IP$_2$).}
\label{ACP}
\end{center}
\end{figure}

The inclusive muon charge asymmetry $a_{CP}$ is also measured in 27 bins:
9 bins of $(p_T, |\eta|)$ $\times$ 3 bins of transverse impact parameter (IP).
The like-sign dimuon charge asymmetry $A_{CP}$ is also measured in 54 bins:
9 bins of $(p_T, |\eta|)$ $\times$ 6 bins of $(\textrm{IP}_1, \textrm{IP}_2)$.
In all cases the asymmetry does not vary significantly with $(p_T, |\eta|)$.
See \cite{dimu_asym4} for full details. Averaging over the 9 bins of
$(p_T, |\eta|)$ we obtain the results summarized in Figures \ref{aCP} and \ref{ACP}.

The measurements of $a_{CP}$ and $A_{CP}$ in bins of IP
are correlated. 
These measurements are in disagreement with
the Standard Model prediction for CP violation in mixing
and interference of $B^0$ and $B^0_s$ by 3.6 standard deviations.
At the present time we do not understand the origin of this 
discrepancy. It could be due to missing Standard Model
contributions, inaccurate predictions of $\Delta \Gamma_d$
or $a^d_\textrm{sl}$ (i.e. the real and imaginary parts of
the matrix element $\Gamma^d_{12}$) due to low energy non-perturbative effects,
an experimental issue that has evaded all cross-checks to 
date, or New Physics. $\Delta \Gamma_d$ is the decay rate difference of the
eigenstates of the $(B^0, \bar{B}^0)$ system, and $a^d_\textrm{sl}$
is the semi-leptonic charge asymmetry of decays of $B^0$
and $\bar{B}^0$ to ``wrong sign" muons.

\begin{figure}
\begin{center}
\scalebox{0.4}
{\includegraphics{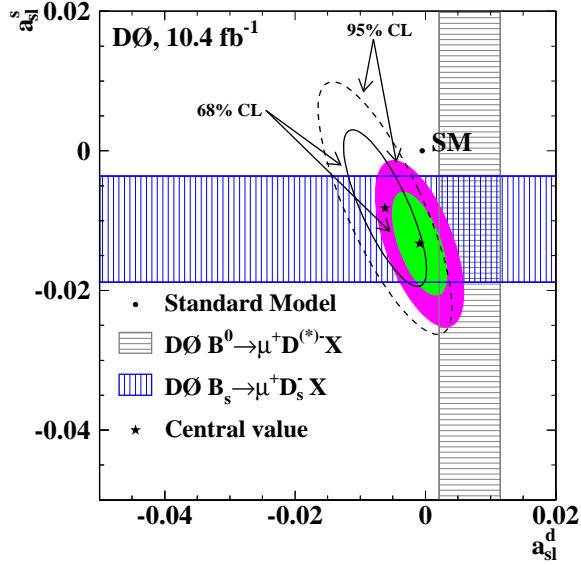}}
\caption{The 68\% and 95\% confidence level 
contours from fit with $\Delta \Gamma_d / \Gamma_d = 0.0050$
(corresponding to the best fit value), and combination of all D\O\ 
measurements (filled areas).
}
\label{ad_as}
\end{center}
\end{figure}

If we assume that the only sources of charge asymmetry are
CP violation in mixing and interference of $B^0$ and $B^0_s$,
we obtain 3-dimensional likelihood contours in the space of
$a^d_\textrm{sl}$, $a^s_\textrm{sl}$ and $\Delta \Gamma_d$.
A slice is shown in Figure \ref{ad_as} at the best fit value of $\Delta \Gamma_d$ 
\cite{dimu_asym4}$^,$ \cite{D0ad}$^,$ \cite{D0as}.

\section{Conclusions}
We obtain a 3.6 standard deviation discrepancy between
the measurements and the Standard Model prediction of
CP violation in mixing and interference of $B^0$ and $B^0_s$.
At the present time we do not understand the origin of this
discrepancy, which has not yet been confirmed nor ruled
out by other experiments, see Table \ref{table1}. 
At D\O\ we are currently measuring
the like-sign dimuon charge asymmetry of sub-samples
of the data set with the hope of solving this puzzle.

\begin{table}
\caption[]{Contributions to $A_S$ allowed by experiments
(with $\pm 1 \sigma$ confidence).
Compare with the measured $A_S = (-0.319 \pm 0.132)\%$.
This is an update of Table II of \cite{CPV_interference}
using the world averages of \cite{PDG13}.
*From the Standard Model prediction and the constraint from
$B^0_s \rightarrow J/\psi \phi$ we expect this contribution to
be negligible.
**These tree level decays are expected to contribute
negligibly to $A_S$.
***This contribution to $A_S$ is proportional to $\Delta \Gamma_d$ and
assumes the Standard Model prediction $\Delta \Gamma_d = (0.42 \pm 0.08)\%$.
If $\Delta \Gamma_d$ turns out to be larger than the prediction due to
low energy non-perturbative effects, then
this $\Delta A_S$ could be sizable.
}
\label{table1}
\vspace{0.4cm}
\begin{center}
\begin{tabular}{|lll|}
\hline
CP violation in & $\Delta A_S$ allowed by exp. & Comments\\
\hline
mixing $B^0 \leftrightarrow \bar{B}^0$ & $(+0.021 \pm 0.083 )\%$ & from exp. $a^d_\textrm{sl}$ \\
mixing $B^0_s \leftrightarrow \bar{B}^0_s$ & $(-0.374 \pm 0.124 )\%$ & from exp. $a^s_\textrm{sl}$ * \\
interference $B^0 \rightarrow (\bar{B}^0) \rightarrow c \bar{c} d \bar{d}$ &
  $(-0.050 \pm 0.012 )\%$ & from SM $\Delta \Gamma_d$ *** \\
interference $B^0_s \rightarrow (\bar{B}^0_s) \rightarrow c \bar{c} s \bar{s}$ &
  $(-0.0009 \pm 0.0003)\%$ & from exp. $\Delta \Gamma_s$ \\
direct decays $b \to c \bar{c} \bar{d}$ & $(+0.003 \pm 0.013)\%$ & from exp. Br \& CPV \\
direct decays $b \to c \bar{c} \bar{s}$ & $(+0.000 \pm 0.043)\%$ & from exp. Br \\
direct decays $b \to \mu X$ & $(-0.17 \pm 0.39 )\%$ & from $a_{CP}$ ** \\
direct decays $c \to \mu X$ & $(-0.07 \pm 0.17 )\%$ & from $a_{CP}$ ** \\
production $p \bar{p} \rightarrow b \bar{c} X$ & $(-0.15 \pm 0.35)\%$ & from $a_{CP}$ \\
neutrinoless double $\beta$ decay & $(-0.12 \pm 0.29)\%$ & from $a_{CP}$ \\
direct decay to ``right sign" $\mu$ & $(-0.13 \pm 0.30 )\%$ & from $a_{CP}$ \\
direct decay to ``wrong sign" $\mu$ & $(-0.46 \pm 1.07 )\%$ & from $a_{CP}$ \\
\hline
\end{tabular}
\end{center}
\end{table}

\end{document}